\documentclass{acm_proc_article-sp}

\usepackage{algorithm}
\usepackage{algorithmicx}
\usepackage{algpseudocode}

\begin{document}
\title{ILCR: Item-based Latent Factors for Sparse Collaborative Retrieval}

\numberofauthors{4} 
%
\author{
%
%
\alignauthor Lu Yu\\
       \affaddr{Alibaba Research Center for Complexity Sciences}\\
       \affaddr{Hangzhou Normal University}\\
       \affaddr{Hangzhou, China}\\
       \email{coolluyu@gmail.com}
\alignauthor Junming Huang\\
	\affaddr{Web Sciences Center}\\
	\affaddr{University of Electronic Science and Technology of China}\\
	\email{mail@junminghuang.com}
\and
\alignauthor Chuang Liu\\
       \affaddr{Alibaba Research Center for Complexity Sciences}\\
       \affaddr{Hangzhou Normal University}\\
       \affaddr{Hangzhou, China}\\
       \email{liuchuang@hznu.edu.cn}
\alignauthor Zi-Ke Zhang\\
       \affaddr{Alibaba Research Center for Complexity Sciences}\\
       \affaddr{Hangzhou Normal University}\\
       \affaddr{Hangzhou, China}\\
       \email{zhangzike@gmail.com}
      }


\maketitle
\begin{abstract}
Interactions between search and recommendation have recently attracted significant attention, and several studies have shown that many potential applications involve with a joint problem of producing recommendations to users with respect to a given query, termed $Collaborative$ $Retrieval$ (CR). Successful algorithms designed for CR should be potentially flexible at dealing with the sparsity challenges since the setup of collaborative retrieval associates with a given $query$ $\times$ $user$ $\times$ $item$ tensor instead of traditional $user$ $\times$ $item$ matrix. Recently, several works are proposed to study CR task from users' perspective. In this paper, we aim to sufficiently explore the sophisticated relationship of each $query$ $\times$ $user$ $\times$ $item$ triple from items' perspective. By integrating item-based collaborative information for this joint task, we present an alternative factorized model that could better evaluate the ranks of those items with sparse information for the given query-user pair. In addition, we suggest to employ a recently proposed scalable ranking learning algorithm, namely BPR, to optimize the state-of-the-art approach, $Latent$ $Collaborative$ $Retrieval$ model, instead of the original learning algorithm. The experimental results on two real-world datasets, (i.e. \emph{Last.fm}, \emph{Yelp}), demonstrate the efficiency and effectiveness of our proposed approach.

\end{abstract}




\section{Introduction}
The task of \emph{Collaborative Retrieval} (or CR for short), producing recommendations to a particular individual with respect to a certain query describing her instant interests, arises from the rapidly emerging needs to explore relative information from huge resource. For example, a user may hope web application to retrieve some songs or movies similar to his/her recent favourites depending on a given query. Collaborative retrieval roots in personalized recommender systems and information retrieval systems but is largely different from a simple combination of those two isolated tasks. Expanding a user-item preference matrix in a recommender system or a query-document relevance matrix in an information retrieval system to a query-user-item affinity tensor, a collaborative retrieval system suffers more severely from data sparsity, especially in the scenario that the explosive growth of information in recent decades has built a huge candidate space.

To tackle the sparsity problem in collaborative retrieval, an effective and efficient method is required to fulfill the $query\times user\times item$ tensor. Tensor factorization models~\cite{rendle2009kdd, symeonidis2008} work well to extend collaborative filtering to recommend items in a tensor setup, but usually designed for special recommendation tasks like tag, mobile app, etc. According to \cite{weston:latent}, most of them can not fully fit the collaborative retrieval task. Additional source is another considerable option to reduce the pain of sparsity, including integrating side information like social networks \cite{ma2011recommender} or trust-based network \cite{jamali}. Nevertheless, the extra noise brought by those approaches severely affect sparse objects.
In the other school, less noisy object relations are leveraged to supplement missing features of sparse objects, such as leveraging similar users~\cite{hsiao2014social} and similar items~\cite{sarwar2001item, itembased2004}. An early exploration to incorporate object relation in collaborative retrieval is the \emph{Latent Collaborative Retrieval} (LCR) model proposed by Weston \emph{et al.}~\cite{weston:latent}. It leverages user-user similarity in modelling a $query\times user\times item$ tensor, in hope of reducing the pain of sparsity. LCR partially addressed the sparsity problem but ignores the item side, which should not be negligible for two reasons. (1) Items suffer from the sparsity problem more severely than users, since items are usually observed with fewer features to support a feature-based or content-based algorithm. (2) Users are dynamic but items are relatively static, which makes user-user similarity less stable and reliable than item-item similarity, as reported in \cite{linden:amazon.com}.

Inspired by hose intuitions, we propose in this paper an \emph{Item-based Local Collaborative Retrieval} (ILCR) model, which leverage item-item similarity to solve the sparsity problem in collaborative retrieval. Specifically, we use a triple $(q, u, a)$ to describe a fact that a user $u$ prefers to an item $a$ with content features correlated to query $q$. Viewing items as media vertices, we utilize item's neighbours with similar tastes to overcome the sparsity problem caused by the lack of content features. 
The ILCR model is trained with a recent proposed pairwise learning algorithm~\emph{Bayesian Personalized Ranking} (BPR)~\cite{rendle2009bpr}, which has been applied in many published recommendation tasks including tag recommendation \cite{rendle2010pairwise}, relation extraction \cite{relationextraction}, focused matrix factorization for advertisement~\cite{focusedmfAd}. As pointed out in~\cite{hsiao2014social, weston2010large} and verified on a real dataset \emph{Last.fm}\footnote{http://www.last.fm}~\cite{cantador2011second}, BPR significantly reduces the runtime required to train a latent model like LCR or ILCR while keeping the same performance, compared with the training strategy \emph{Weighted Approximate-Rank Pairwise} (WARP) that was previously used in LCR.

In summary, the main contributions of this paper include:

\begin{itemize}
  \item Comparing with WARP learning algorithm, the application of BPR algorithm on optimizing the parameters of LCR can sharply increase the training speed and simultaneously preserve almost the same performance on evaluation metric.
  \item We consider the item-based collaborative information to sufficiently alleviate the sparsity problem caused by items' lack of content features, and subsequently propose the ILCR method.
  \item The experimental results on the \emph{Last.fm} and \emph{Yelp} datasets show that the proposed algorithm ILCR is superior to LCR model, especially, when dealing with a dataset containing a massive amount of items with sparse information.
\end{itemize}

\section{Collaborative Retrieval}
The objective of CR can be simply defined as generating a personalized ranking list of items to fit a particular user's tastes with respect to a given query. To achieve this goal, the proposed approaches should clearly define a scoring function $f$ to represent the relevance of a given triple (query,user,item) $\in \mathcal{Q}\times\mathcal{U}\times\mathcal{A}$, where $\mathcal{Q}$, $\mathcal{U}$, $\mathcal{A}$ denote the set of queries, users, items, respectively. In practice, only the top-$k$ retrieved items could draw users' attention. Thereby, the learned scoring function $f$ should promote users' interesting items to high position as much as possible for a particular query. Generally, the parameters of $f$ can be derived from the training samples by optimizing a pre-defined ranking loss function.

We next briefly review the aspects of LCR model for CR task. Then, we detail the principles of the proposed model, namely ILCR.

\subsection{Latent Collaborative Retrieval}
The central idea of latent collaborative retrieval (LCR) is to represent each entity (e.g. user, query, item) as a $n$-dimensional feature vector. Analogous to MF-based approaches \cite{koren2009matrix}, the relationship between each pair entities can be measured by calculating the dot-product of them. Formally, LCR's parameter space includes matrices $S\in \mathbb{R}^{|\mathcal{Q}|\times n}$, $V\in \mathbb{R}^{|\mathcal{U}|\times n}$, and $T\in \mathbb{R}^{|\mathcal{A}|\times n}$, which denotes the feature matrix of queries, users, and items, respectively. To preciously evaluate a user's preference on a item w.r.t. a given query, LCR additionally allocates each user $u$ an encoder matrix $U_{u}\in \mathbb{R}^{n \times n}$. The scoring function $f$ of LCR model can be then given as follows:

\begin{equation}
    \label{LCRF}
    f(q,u,a) = S_{q}U_uT_{a}^\top + V_{u}T_{a}^\top,
\end{equation}
where $S_q$ represents the row of $S$ corresponding to query $q$, $V_u$ is the row of $V$ corresponding to user $u$, and $T_a$ denotes the row of $T$ corresponding to item $a$. Intuitively, the first term in Equation (\ref{LCRF}) distinguishes LCR from the proposed tensor models for context-aware CF\footnote{The goal of context-aware CF is to produce personalized recommendations to a particular user by taking into account the contextual information, such as time, location, and so on. Therefore, the query could also be regarded as a context factor.} \cite{adomavicius2011context, karatzoglou2010multiverse,rendle2010pairwise,baltrunas2011matrix} tasks. The second term independent of the query denotes users' basic preferences on items.

\section{Item-based Latent Collaborative Retrieval}
In this section, we will firstly describe our motivation behind the idea of Item-based LCR. Then we will show how to measure the relevance of $(query, user, item)$ triple from item-based perspective after graphically revisiting LCR method.
\subsection{Motivation}
In real applications, we often suffer from the problem of $asymmetric$ $information$, that is to say, characteristics of a majority of items can not be fully expressed due to the lack of specific description. By contrast, massive amount of popular items are represented as adequate descriptive terms. In this circumstance, the retrieval systems would unfairly evaluate the ranks of those items with sparse information since current description can not cover the characteristics of them. In order to cope with this problem, many web applications, such as $delicious$\footnote{https://delicious.com/}, douban\footnote{http://www.douban.com}, integrate the function of $collaborative$ $tagging$ \cite{jaschke2007tag}, which offers privileges for users to add free-style keywords to the shared content, e.g. books, movies, videos. To a certain extent, collaborative tagging enriches the content of items with descriptive terms for future filtering or search. However, there are still a huge amount of items with sparse information because of the lack of motivated users to share their tags. Therefore, the dataset that we have deeply involved mainly contains items either represented as adequate keywords, or oppositely with seldom keyword.
\begin{figure*}
\centering
\begin{tabular}{ccc}
\epsfig{file=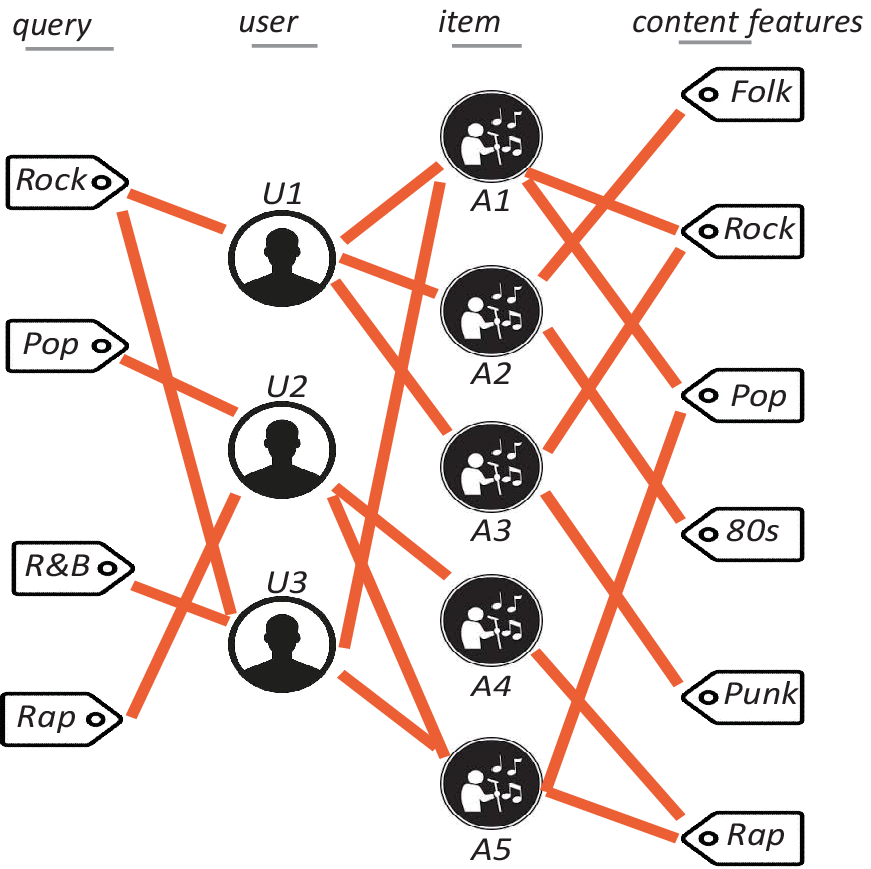, scale = 0.65} \hspace{1.2cm}&
\epsfig{file = 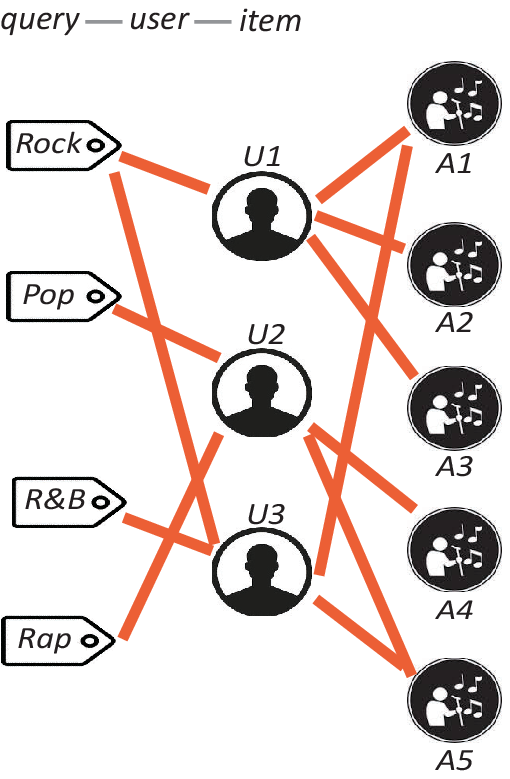, scale = 0.73} \hspace{0.3cm} &
\epsfig{file = 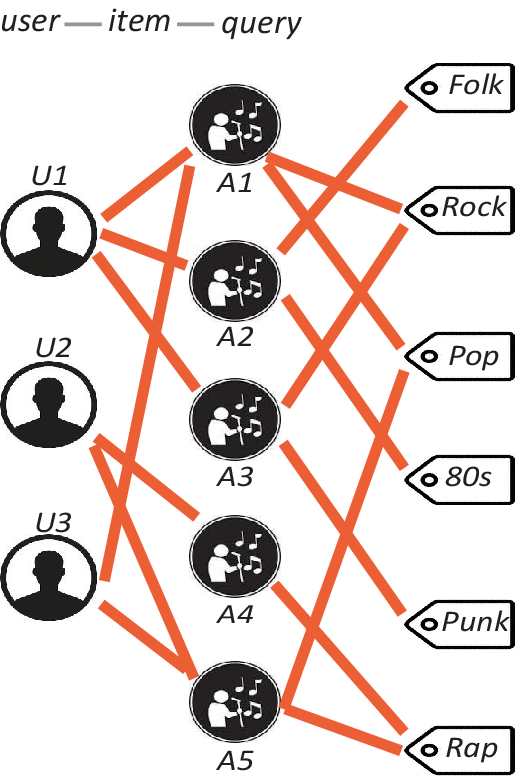, scale = 0.73}\\
(a) & (b) & (c)
\end{tabular}
\caption{Graphical representation of the $user$ $\times$ $query$ $\times$ $item$ relationship.}
\end{figure*}

To promote the performance of retrieval systems, we assume that the characteristics of those items with sparse information, denoted as $sparse$ $items$, can be derived from their neighbours with abundant descriptive terms. Typically, one can firstly employ efficient tag recommendation approaches \cite{garg2008personalized, jaschke2007tag} to pre-target sparse items with most possible keywords or tags. Then, retrieval systems could return a ranked list of items by computing the scores for all items w.r.t. a given user-query pair. However, the pre-targeting process could cost lots of computation resources. This significant challenge impulses us to think ``Can we invent some algorithms to naturally represent the ternary interaction of  $(query, user, item)$ in CR task?"

To answer this question, we next graphically review the relationship of  $(user, query, item)$ and interpret how to model the ternary relationship from both users' and items' perspective.

\subsection{Revisit Ternary Relationship}
To formulate our ideal, we firstly represent the relationship of the $user$ $\times$ $query$ $\times$ $item$ triple as a graph, simply shown in Figure 1.(a), where entities $(user, query, item)$ are represented as vertices and red edges represent the observed relationship between a pair of entities. In order to intuitively describe the ternary relationship of $(user, query, item)$, we decompose Figure 1.(a) into two parts showed in Figure 1.(b) and Figure 1.(c), respectively. In Figure 1.(b), each positive observation $(user,query,item) \in \textbf{X}$ can be represented as a route path $query$-$user$-$item$\footnote{Here we use ``path" to distinguish the representation of $(user, query, item)$ relationship from different perspectives.}, where users serve as the intermedia vertices responsible for transforming the ''resource" between items and queries, and the bipartite graph in the right intuitively represent the users' preferences over items. In terms of Figure 1.(c), the relationship of each positive observation is represented as a route path $user$-$item$-$query$, where differently items serve as the intermedia ''resource" transformer between users and queries, and the left bipartite graph represent the rating information of user-item pairs. In the following part, we will give insight into the possible effects brought by the slight difference between Figure 1.(b) and (c). 

It can be seen that Figure 1.(b) illustrates an user-central behaviour network, where the relationship of user-item pairs is represented as a collaborative network in the right part, analogously, the left part depicts the collaborative interactions of user-query pairs. In practical application, Figure 1.(b) could be extremely sparse due to the tinily available querying or rating information from users. In order to deal with such challenge caused by the lack of users' historical behaviours, the proposed approach for CR task should combine both collaborative networks in Figure 1.(b). By doing this, users' preferences over unconnected queries or items could be induced from their neighbours that have similar querying bias, or tend to rate a similar set of items.

Turning to Figure 1.(c), we can easily find that left part is shared with Figure 1.(b), but it represents a collaborative network from items' perspective. Analogous to Figure 1.(b), the right part of Figure 1.(c) represents the collaborative network of item-query pairs. Since content features of items might be the possible queries, the word ``query" in Figure 1.(c) generally indicate both content features and the observed queries. With the problem of $asymmetric$ $information$ mentioned in Section 3.1, massive amount of items in Figure 1.(c) actually have too limited content to represent intrinsic feature. To better measure the ternary relationship of $(user, query, item)$, this challenge should be taken into account when we attempt to design an innovative algorithm for CR task. In CF field, many benefits of item-based methods have been fully discussed, among of which using the characteristics of neighbours with similar features to induce the target item's unknown features might be the possible way to help us to better evaluate the rank of sparse items w.r.t. a given user-query pair. However, the case in CR task is totally different from traditional CF problem, which usually involves binary relationship between user-item pairs, rather than the ternary interactions among $(user, query, item)$. As we mentioned in Section 3.1, pre-targeting items could cost a lot of computation in searching similar items. In this work, we prefer to combine both user-item and item-query collaborative networks to better capture the latent relationship of $user$-$item$-$query$. We believe that it could further tackle the sparsity problem caused by the lack of content features, meanwhile improve the effect of learning the ternary relevance of $user$-$item$-$query$ path.

\subsection{Our Method}
As mentioned above, we consider items' collaborative information based on the assumption that regarding items as the media vertex could leverage the similar items \cite{sarwar2001item} with rich descriptive terms to improve the performance of retrieval systems on estimating the ranks of sparse items. In our model, the score of $user$-$item$-$query$ path in Figure 1.(c) can be given by:
\begin{equation}
    \label{QAU}
    g(q,a,u) = S_qA_aV_u^\top,
\end{equation}
where $A_a \in \mathbb{R}^{n\times n}$ is the linear transformation matrix of item $a$. Finally, we propose to solve CR task by integrating the item-based collaborative information into LCR model, namely $Item$-$based$ $latent$ $collaborative$ $retrieval$ (ILCR). Consequently, Equation (1) can be modified as:
\begin{equation}
    \label{LCRwQP}
    \begin{aligned}
    f(q,u,a) &= S_{q}U_uT_a^\top + V_{u}T_a^\top + S_qA_aV_u^\top
    \end{aligned}
\end{equation}
where the first term intuitively captures the ternary interaction of $(query, user, item)$ by using encoder matrix $U_u$ of each user to indirectly combine both user-query and user-item collaborative networks into one model. The querying and rating preferences of target users could be learned based on not only their own historical behaviours, but also their neighbours'. To some extent, it could effectively deal with the sparsity problem caused by the lack of users' historical information. Differently, the third term in Equation (\ref{LCRwQP}) focuses on modelling items' collaborative information to tackle the sparsity problem caused by the items' content features. 

In terms of the scoring function, one might think that ILCR is slightly different from LCR, i.e. the performance of ILCR could be almost the same with LCR.  But actually, the experimental results in Section 4.4 show that ILCR can better estimate the ranks of items w.r.t. a given user-query pair after considering the item-based collaborative information. The reason is that ILCR explicitly models item-based interactions in a sophisticated system that might be difficult to find for LCR method.

\subsection{Pairwise Learning Approaches}
The unknown parameters of the scoring function should be efficiently learned under the assumption that the top $k$ retrieved items should include as many profitable selections as possible for a particular user w.r.t a given query. To achieve this, the CR task can be generally formulated as generating a ranked list of items for a given ($u$, $q$) pair by solving a pairwise ranking problem. A common approach is to regard each observation in a given training set $\textbf{X}$, including $m$ observations ${(q_i,u_i,a_i)}_{i=1,2,\ldots,m} \in \mathcal{Q}\times\mathcal{U}\times\mathcal{A}$, as a positive retrieval event, otherwise as a negative example. Subsequently, the selected learning algorithm should give the definition of $pairwise$ $violation$ cost, which would be produced if a negative item is assigned with a larger score or within a "margin" from the positive item.

In this section, we shall firstly detail the mechanism of WARP, originally used in the first piece of CR work \cite{weston:latent}, then introduce a generic learning algorithm proposed by Rendle et al. \cite{rendle2009bpr}, namely BPR. Finally, we will show how to adapt BPR instead of WARP to correctly learn the $f$ function of LCR and our model, ILCR. 

\subsubsection{Weighted Approximate-Ranking Pairwise (WARP)}
The WARP Loss can be defined as follows:
\begin{equation}
    \label{WARP}
    err_{WARP} = \sum_{i=1}^mL(rank_{a_i}(\bar{f}(q_i,u_i))),
\end{equation}
where $\bar{f}(q_i,u_i)$ is a vector, which contains predictions for all items for a fixed user-query pair. The $a_{i}^{th}$ element of $\bar{f}(q_i,u_i)$, denoted as $\bar{f}_{a_i}(q_i,u_i)$, is the predicted relevance value of $i^{th}$ training example $(q_i,u_i,a_i)$. Correspondingly, $rank_{a_i}(\bar{f}(q_i,u_i))$ in Equation (\ref{WARP}) is a margin-based rank of item $a_i$, defined as:
\begin{equation}
    \label{RANK}
    rank_{a_i}(\bar{f}(q_i,u_i)) = \sum_{b\neq a_i} \mathbb{I}[1 + \bar{f}_b(q_i,u_i) \geq \bar{f}_{a_i}(q_i,u_i)],
\end{equation}
where $\mathbb{I}[\cdot]$ is the indicator function. In Equation (\ref{WARP}), $L$ is a weight function evaluating the loss of the current scoring function $f$:
\begin{equation}
    \label{L}
    L(k) = \sum_{i=1}^{k} \alpha_i,\ with\ \alpha_1 \geq \alpha_2 \geq \alpha_3 \geq \ldots \geq0.
\end{equation}
According to the advices of \cite{weston:latent}, we choose $\alpha_i = 1/i$ as the weighting approach, which would assign large weights to top positions with rapidly decaying weight for lower positions. Intuitively, optimizing the WARP loss means to rank each positive item $a_i$ in the training set to highest position. For example, given a random query-user pair denoted as $(q_i, u_i)$, if the score of an uncollected item $b$ is less than a margin of one from the score of $a_i$, this pair will produce a cost.

Subsequently, Equation (\ref{WARP}) could be optimized by gradient descent based algorithms. However, in each updating step, it is expensive to compute the exact value of $rank_{a_i}$ for each observation when the number of items is very large. Thus, the exact rank of Equation (\ref{RANK}) can be estimated by a random sampling process at each step \cite{weston:latent}. That is to say, for a given observed sample $(q_i, u_i, a_i)$, one uniformly draws at random items from $\mathcal{A}$ until finding a violated item $b$, which satisfies $1\ +\ f(q_i,\ u_i,\ b) > f(q_i,\ u_i,\ a_i)$, and then the rank of $a_i$ can be approximated as
\begin{equation}
    \label{APRANK}
    rank_{a_i}(\bar{f}(q_i, u_i)) \approx \lfloor \frac {|\mathcal{A}| - 1} {K} \rfloor,
\end{equation}
where $\lfloor \cdot \rfloor$ is the floor function and $K$ is the number of steps needed to find a item $b$. Then Equation (\ref{WARP}) can be modified as
\begin{equation}
    \label{MWARP}
    \begin{aligned}
    &\overline{err_{WARP}} = \sum_{i=1}^mL_i\\
    &L_i = L(rank_{a_i}(\bar{f}(q_i,u_i))) \cdot |1 - f(q_i, u_i, a_i) + f(q_i, u_i, b)|.
    \end{aligned}
\end{equation}

 Finally, Equation (\ref{MWARP}) could be optimized by stochastic gradient descent (SGD). We constrain the parameters using $||S_i|| \leq C$, $i\in \{1,\ldots,|\mathcal{Q}|\}$, $||V_i|| \leq C$, $i\in \{1,\ldots,|\mathcal{U}|\}$, $||T_i|| \leq C$, $i \in \{1,\ldots,|\mathcal{U}|\}$ and project the parameters back into the constraints at each SGD step.
 
\subsubsection{Bayesian Personalized Ranking (BPR)}
 To clearly describe BPR algorithm \cite{rendle2009bpr}, we leverage a notation $D_X$ to denote the pairwise ranking constraints.
\begin{equation*}
    \label{DX}
    D_X = \{(u,q,a,b):(u,q,a)\in \textbf{X} \wedge (u,q,b) \not\in \textbf{X}\}
\end{equation*}
Next, we present a generic approach to solve the personalized ranking problem for CR tasks by maximizing the following posterior probability of the parameter space $\Theta$.
\begin{equation}
    \label{POSBPR}
    p(\Theta|>_{u,q}) \varpropto p(>_{u,q}| \Theta)p(\Theta),
\end{equation}
\begin{figure}
\centering
\epsfig{file=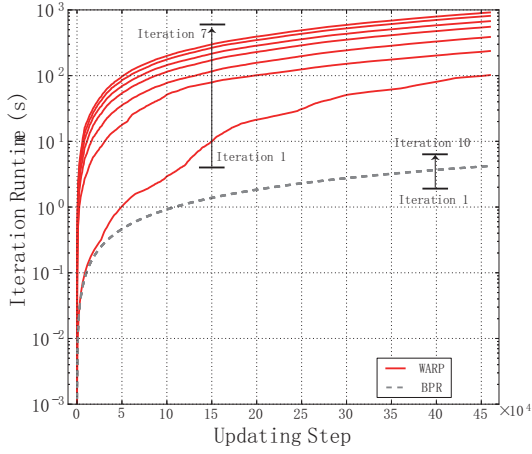, height=6cm, width=7cm}
\caption{Accumulated runtime required by WARP and BPR for searching violated items in each training iteration. $x$ axis denotes the number of updating steps in a training iteration. $y$ axis denotes the accumulated runtime.}
\end{figure}
where $p(\Theta)$ denotes the prior probability of $\Theta$, and notation $>_{u,q} = \{a >_{u,q} b: (u,q,a)\in \textbf{X},(u,q,b) \not\in \textbf{X}\}$ denotes the pairwise ranking structure for a given $(u,q)$ pair. We assume that each element of $>_{u,q}$ is independently drawn from the same probability. Hence, the above likelihood function $p(>_{u,q}| \Theta)$ can be rewritten as:
\begin{equation}
    \label{LIKELIHOOD}
    p(>_{u,q}| \Theta) = \prod_{(u,q,a,b)\in D_X}p(a >_{u,q} b | \Theta),
\end{equation}
where $p(a >_{u,q} b | \Theta)$ denotes the probability that a user really prefers item $a$ to item $b$ for a given query, defined as \cite{rendle2009bpr}:
\begin{equation}
    \label{LOGISTICS}
    \begin{aligned}
    p(a >_{u,q} b | \Theta) &= \sigma(\hat{x}_{u,q,a,b}(\Theta))\\
    \sigma(x) &= \frac {1} {1 + e^{-x}}.
    \end{aligned}
\end{equation}
Here we choose $\hat{x}_{u,q,a,b}(\Theta) = \hat{x}_{u,q,a} - \hat{x}_{u,q,b}$, in fact $\hat{x}_{u,q,a,b}(\Theta)$ could be a arbitrary real-valued function depending on the parameters $\Theta$. For $p(\Theta)$, we define it as a normal distribution with zero mean and covariance matrix $\sum_{\Theta} = \lambda_{\Theta}I$, that is, $\Theta \backsim \mathcal{N}(\textbf{0},\sum_{\Theta})$. Now we can infer the BPR-$\textsc{Opt}$ by filling $p(\Theta)$ into the maximum posterior probability in Equation (\ref{POSBPR}).
\begin{equation*}
    \label{RPOSBPR}
    \begin{aligned}
    BPR\text{-}\textsc{Opt} &= \ln p(\Theta|>_{u,q}) \\
    &= \ln \prod_{(u,q,a,b)\in D_X}p(a >_{u,q} b | \Theta)p(\Theta)\\
    &= \ln \prod_{(u,q,a,b)\in D_X}\sigma(\hat{x}_{u,q,a,b}(\Theta))p(\Theta)\\
    &= \sum_{(u,q,a,b)\in D_X} \ln\sigma(\hat{x}_{u,q,a,b}) + \ln p(\Theta)\\
    &= \sum_{(u,q,a,b)\in D_X} \ln\sigma(\hat{x}_{u,q,a,b}) - \lambda_{\Theta}||\Theta||^2,
    \end{aligned}
\end{equation*}
where $\lambda_{\Theta}$ are model regularization parameters.
\begin{figure}
\centering
\epsfig{file=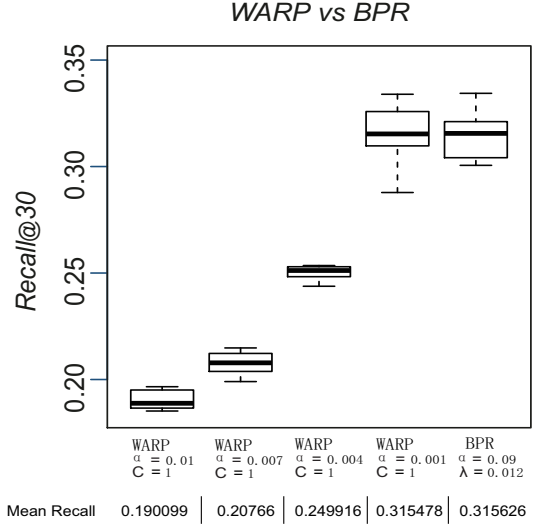, height = 7cm, width = 7cm}
\caption{Comparisons of WARP and BPR on recall@$30$ (see Section 3.2). Four groups of (Learning rate $\alpha$, constraint $C$) setups are randomly selected for WARP. One group of (Learning rate $\alpha$, regularization $\lambda$) setup chosen for BPR mimics the best performance of WARP. The performance of each learning algorithm is validated on the same testing dataset (see Section 3.1).}
\end{figure}

Typically, gradient ascent based algorithm is an apparent optimization strategy for maximizing the posterior probability in Equation (\ref{POSBPR}). However, standard gradient ascent is not suitable for BPR-$\textsc{Opt}$ because we have to compute $\hat{x}_{u,q,a,b}$ for all negative items $b$ w.r.t. a given training sample $(u,q,a) \in \textbf{X}$. If we have a large amount of items, it will be inefficient to update the parameters in each gradient ascent step. To effectively learn the parameters, a stochastic gradient-ascent (SGA) based algorithm, namely $\textsc{LearnBPR}$ , was proposed by Rendle \emph{et\ al.} \cite{rendle2009bpr} for optimizing BPR-$\textsc{Opt}$. Instead of comparing with all negative items, $\textsc{LearnBPR}$ only draws at random a negative item $b$ in each SGA step, which makes $BPR$ superior to $WARP$ in terms of the computation complexity (see next section). The procedure of $BPR$ is presented in $\textbf{Algorithm 1}$.
\begin{algorithm}
\caption{Optimizing models for BPR with $\textsc{LearnBPR}$}
    \begin{algorithmic}[1]
        \Require $\textbf{X}$: Training Set
        \Ensure Learned $\hat{\Theta}$
        \State initialize $\Theta$
        \Repeat
             \State randomly draw (u,q,a) from $\textbf{X}$
             \State randomly draw (u,q,b) from $(u\times q \times \mathcal{A}) \setminus \textbf{X}$
             \State $\hat{x}_{u,q,a,b} \leftarrow \hat{x}_{u,q,a} - \hat{x}_{u,q,b}$
             \State $\Theta \leftarrow \Theta + \alpha ((1 - \sigma(\hat{x}_{u,q,a,b})) \frac {\partial} {\partial \Theta} \hat{x}_{u,q,a,b} - \lambda_{\Theta}\Theta)$
        \Until convergence.
    \end{algorithmic}
\end{algorithm}

\subsection{Learning LCR and ILCR with BPR}
In above sections, we intuitively review two optional learning algorithms for LCR model. In this section, we will present the comparative results to study the efficiency of both learning algorithms.

In terms of WARP, the random sampling procedure dominates the runtime according to several recent works \cite{weston2010large, hsiao2014social}. Ref. \cite{weston2010large} indicated that each SGD step requires less than $1 + min(\frac {|\mathcal{A}| - 1} {rank_{a}(f(u,q))})$ sampling times to find a violated item $b$ on average. Meanwhile, the computations of scores on items $b$ could increase sharply or even worse for a massive item database, if the positive item $a$ happens to rank at the top of the list. Interestingly, such occasional situation seems to happen frequently \cite{hsiao2014social}. Comparatively, BPR only samples one time at each parameter updating step. In this paper, we conduct several experiments with the same hardware condition to demonstrate the efficiency of BPR for optimizing the LCR model. To avoid the fluctuations coming from parameters randomly initialization on the results, each experiment for Figure 3 runs ten independent times. Experimental results show that BPR needs less runtime to learn the parameters of LCR than WARP (see Figure 2), meanwhile, without decreasing the performance of LCR on the evaluation metric (see Figure 3). In Figure 2, the iteration runtime curves of BPR are difficult to be distinguished from each other, thus runtime curves of ten training iterations are crowded together.

In summary, we propose to employ BPR instead of WARP to optimize LCR model. The complete BPR learning procedure for LCR can be found in $\textbf{Algorithm 2}$, where the function of 5-12th steps is to update corresponding parameters based on the 5-6th steps in $\textbf{Algorithm 1}$.
\begin{algorithm}
\caption{Optimizing LCR with BPR}
    \begin{algorithmic}[1]
        \Require $\textbf{X}$: Training Set
        \Ensure $S,U,V,T$
        \State Initialize $S,U,V,T$ from uniform probability $\mathcal{U}(x, y)$
        \Repeat
             \For {(u,q,a) in $\textbf{X}$}
                 \State randomly draw (u,q,b) from $(u\times q \times \mathcal{A}) \setminus \textbf{X}$
                 \State $\hat{x}_{u,q,a,b} \leftarrow f(u,q,a) - f(u,q,b)$
                 \State Updating $S_q,U_u,V_u,T_a,T_b$
                 \State $loss \leftarrow 1 - \sigma(\hat{x}_{u,q,a,b})$
                 \State $S_q \leftarrow S_q + \alpha (loss \cdot U_u(T_a - T_b)^{\top} - \lambda S_q )$
                 \State $V_u \leftarrow V_u + \alpha (loss \cdot (T_a - T_b) - \lambda V_u)$
                 \State $T_a \leftarrow T_a + \alpha (loss \cdot (S_qU_u + V_u) - \lambda T_a)$
                 \State $T_b \leftarrow T_b - \alpha (loss \cdot (S_qU_u + V_u) + \lambda T_b)$
                 \State $U_u \leftarrow U_u + \alpha (loss \cdot S_{q}^\top (T_a - T_b) - \lambda U_u)$
             \EndFor
        \Until validation performance does not improve.
    \end{algorithmic}
\end{algorithm}

As Eequation (\ref{LCRwQP}) shows, the model parameters of ILCR include:
\begin{equation*}
\begin{aligned}
T \in &\mathbb{R}^{|\mathcal{A}| \times n}, V \in \mathbb{R}^{|\mathcal{V}| \times n}, S \in \mathbb{R}^{|\mathcal{Q}| \times n}\\
 &A \in \mathbb{R}^{|\mathcal{A}| \times n \times n}, U \in \mathbb{R}^{|\mathcal{V}| \times n \times n}
\end{aligned}
\end{equation*}

For a given training dataset, the values of parameters in ILCR can be learned by following learning procedure of BPR.  The complete learning procedure of ILCR is described in $\textbf{Algorithm\ 3}$, where the function of 5-14th steps is to update corresponding parameters based on the 5-6th steps in $\textbf{Algorithm 1}$.
\begin{table}
\centering
\caption{Basic statistics of $Lastfm$ and $Yelp$ dataset.}
\begin{tabular}{cccccc} \hline
dataset & user & item & query & samples & sparsity\\ \hline
lastfm-50tags & 1,529 & 8,669 & 50 & 574,521 & 99.91\% \\ \hline
Yelp & 16,826 & 14,902 & 587 & 806,261 & 99.99\% \\ \hline
\end{tabular}
\end{table}
\begin{algorithm}
\caption{Optimizing ILCR with BPR}
    \begin{algorithmic}[1] 
        \Require $\textbf{X}$: Training Set
        \Ensure $S,V,U,T,A$
        \State Initialize $S,V,U,T,A$ from uniform probability $\mathcal{U}(x, y)$
        \Repeat
             \For { $(u,q,a)$ in $\textbf{X}$}
                 \State randomly draw (u,q,b) from $(u\times q \times \mathcal{A}) \setminus \textbf{X}$
                 \State $\hat{x}_{u,q,a,b} \leftarrow f(u,q,a) - f(u,q,b)$
                 \State Updating $S_q$,$V_u$,$T_a$,$T_b$,$U_u$,$A_a$,$A_b$
                 \State $loss \leftarrow 1 - \sigma(\hat{x}_{u,q,a,b})$

                 \State $S_q \leftarrow S_q + \alpha (loss \cdot (U_u(T_a - T_b)^{\top} + (A_a - A_b)V_u^{\top}) - \lambda S_q )$
                 \State $V_u \leftarrow V_u + \alpha (loss \cdot (T_a - T_b + S_q(A_a - A_b)) - \lambda V_u)$
                 \State $T_a \leftarrow T_a + \alpha (loss \cdot (S_qU_u + V_u) - \lambda T_a)$
                 \State $T_b \leftarrow T_b - \alpha (loss \cdot (S_qU_u + V_u) + \lambda T_b)$
                 \State $U_u \leftarrow U_u + \alpha (loss \cdot S_{q}^\top (T_a - T_b) - \lambda U_u)$
                 \State $A_a \leftarrow A_a + \alpha (loss \cdot S_{q}^\top V_u - \lambda A_a)$
                 \State $A_b \leftarrow A_b - \alpha (loss \cdot S_{q}^\top V_u + \lambda A_b)$
             \EndFor
        \Until validation performance does not improve.
    \end{algorithmic}
\end{algorithm}

\section{Experimental Results}
In this section, we present experimental results to demonstrate the efficiency of the proposed collaborative retrieval model (ILCR) on two real-world datasets, one of which is obtained from the Last.fm music website, the other one is from $Yelp$ $Dataset$ $Challenge$\footnote{http://www.yelp.com/dataset\_challenge} Round 3.

\subsection{Dataset \& Preprocessing}

$Last.fm$ dataset is released in the framework of HetRec 2011, called $hetrec2011$-$lastfm$-$2k$ ($lastfm$-$2k$) \cite{cantador2011second}, and contains heterogeneous information, which mainly covers users' collaborative tagging behaviours, listening preferences on artists, as well as social relationship.

In this paper, we focus on resources associated with collaborative retrieval tasks, that is, users' listening and tagging logs, which contain more than ten thousand unique tags utilized by 1892 users to reveal the characteristics of 17632 listened artists. While, these tags are used to represent artists with personal "notes" rather than widely accepted genres. To remove this noise, we keep only the top 50 most used tags, generally correlated to the genres of music: for example, the top 5 tags are "rock", "pop", "alternative", "electronic", and "indie".

To implement CR task, we need preprocess the dataset. It's noted that the query analysis applications will transform the provided queries as keywords or phrases which are the basic units to represent the items. Then retrieval system return a ranked list of items based on the parsed queries. Thereby, a common approach is to regard a $query$$\times$$user$$\times$$item$ triple as a $keyword$$\times$$user$$\times$$item$, where the keywords are equivalent to the set of filtered user tags, or content features. We carry out data preprocessing with the following steps to obtain an expectant dataset, denoted as $lastfm$-$50tags$, for the comparison experiments.

If an artist $a$ has ever been listened and assigned with several tags by a user $u$, for example, $rock$ and $indie$, then we allocate ($rock$, $u$, $a$) and ($indie$, $u$, $a$) to $lastfm$-$50tags$. If the artist is not assigned with any tags by the user $u$, the genres, assigned by other users to $a$, are distributed to the ($u$, $a$) pairs. If no user has ever assigned any genre to $a$, we choose to drop corresponding ($u$, $a$) pairs in default, since such user-artist pairs are not perfect training examples for the CR task. In the end, we infer the $lastfm$-$50tags$ (see Table 1) with 574,521 data points of the form ($u$, $q$, $a$) from the $last.fm$-$2k$ dataset. 

In addition to $last.fm$-$2k$ dataset, this research is also performed on a recent academic dataset, published under the licence of $Yelp$ $Dataset$ $Challenge$ Round 3, hereafter referred to as $Yelp$. This dataset contains 335,022 reviews and ratings given by 70746 users to 15470 businesses located in the Phoenix and AZ metropolitan area. Each business is characterized with rich content, for example, category like ``Restaurants", ``Shopping", ``Health \& Medical", which makes us accessible to have insight into users' preferences over different businesses. We pre-filter this data to contain users with at least 4 reviews, also corresponding businesses. After pre-filtering, we obtain a processed dataset including 806,261 well-formed points with format (user, category, business), given by 16826 users to 14902 businesses, which totally are pre-tagged with 587 business categories. For example, top-10 categories are ``Restaurants", ``Shopping", ``Food", ``Beauty \& Spas", ``Automotive", ``Mexican", ``Health \& Medical", ``Home Services", ``Nightlife", ``Fashion". Each sample (user, category, business) indicates that a particular user ever rated a business associated with the target category. Table 1 summarizes the basic information of the observed datasets.

\begin{figure*}
    \centering
    \begin{tabular}{ccc}
    \epsfig{file=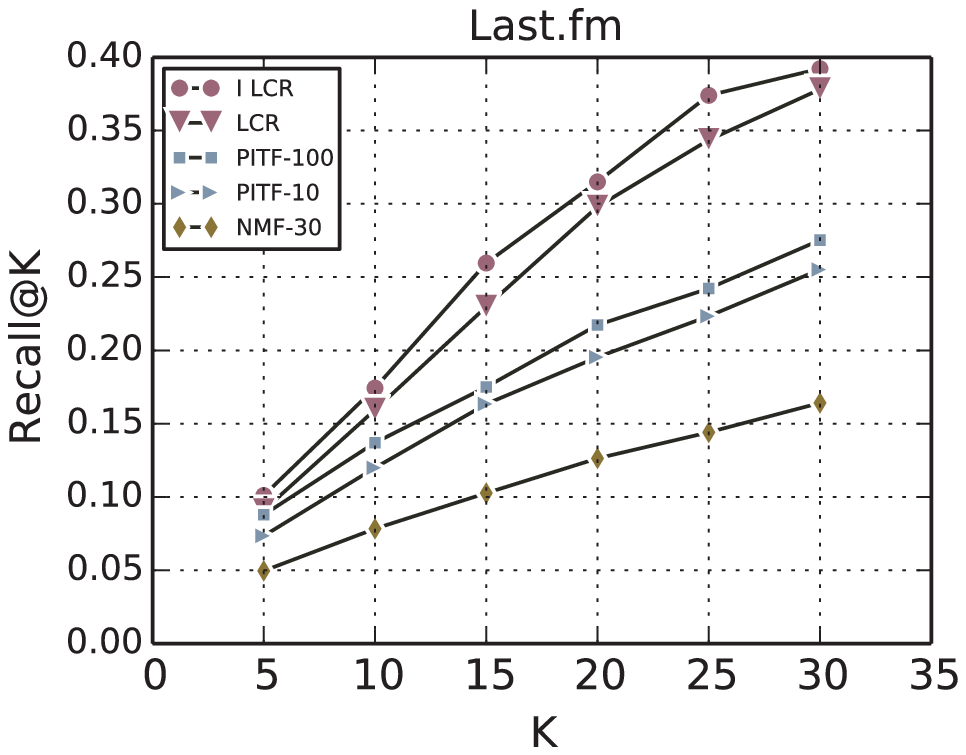, height=5.5cm, width = 5.6cm} & 
    \epsfig{file=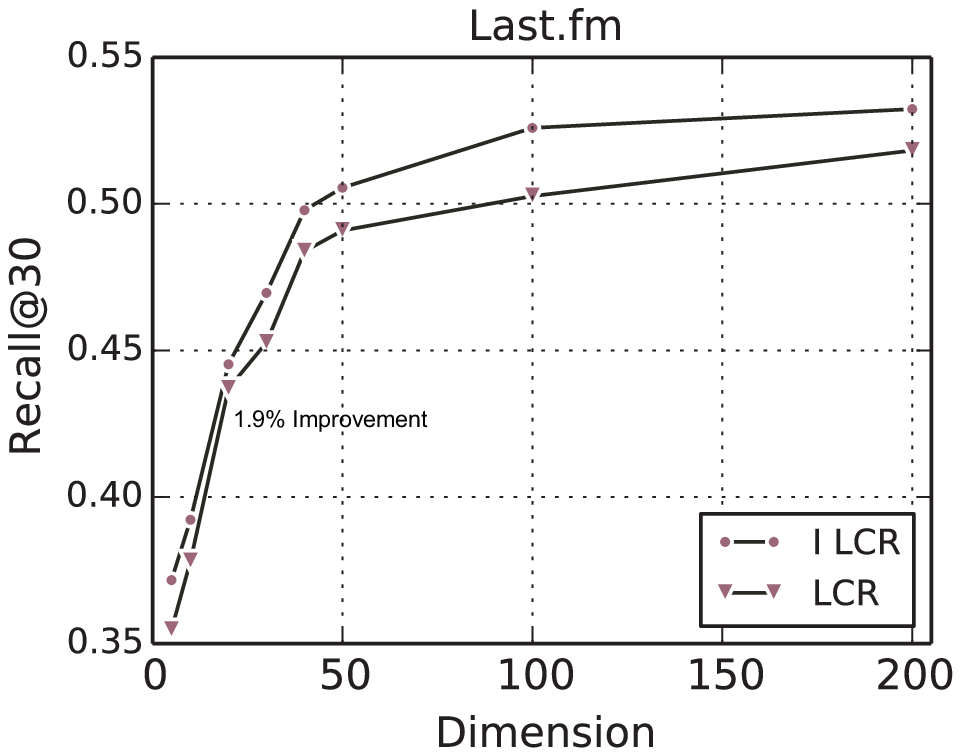, height=5.5cm, width = 5.6cm} &
    \epsfig{file=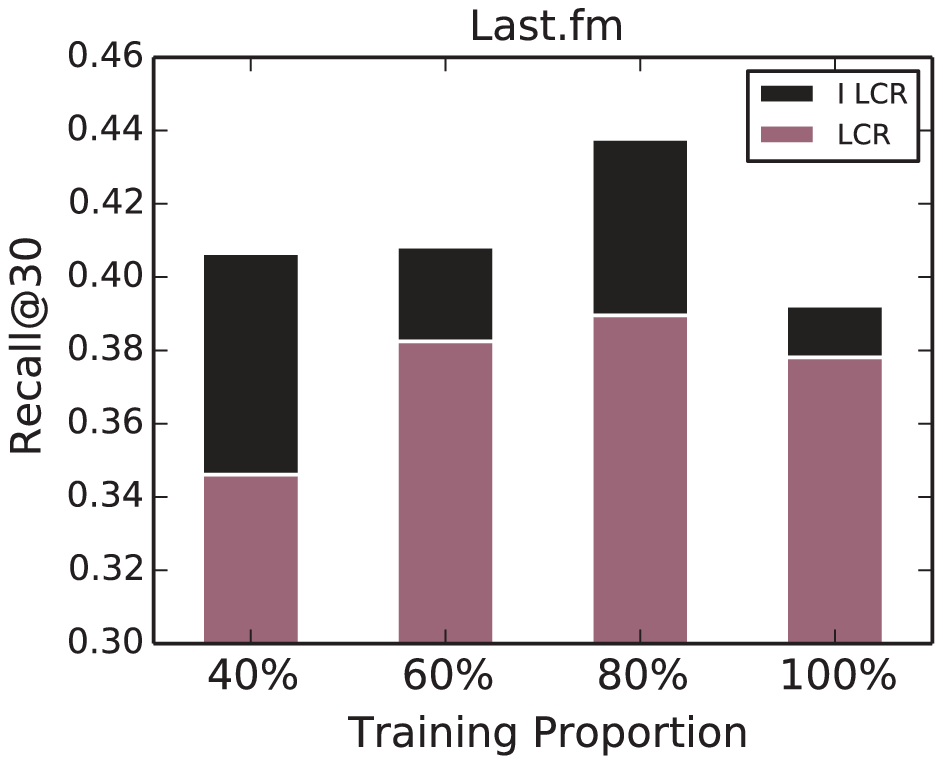, height=5.5cm, width = 5.6cm}\\
    (a) & (b) & (c)
    \end{tabular}
    \caption{Experimental results on $Last.fm$ dataset. (a) Different algorithms' \emph{Recall} vs. different $k$. The integer numbers of PITF and NMF represent the dimension $n$, and $n$ is set as 10 for ILCR and LCR algorithms. (b) Recall@30 vs. the dimension parameter $n$. (c) Recall@30 vs. the dimension parameter $n$.}
\end{figure*}

It's noted that the sparsity of $lastfm$-$50tags$ is 99.91332\%, approximately over 158 times denser than $Yelp$ dataset, which offers available stuff to distinguish the performance of our proposed model from other baseline algorithms in dealing with sparsity problem.

\subsection{Experiment Settings}
To demonstrate the efficiency of our proposed approach, we mainly compare ILCR with the state-of-the-art collaborative retrieval algorithm, that is, LCR model as well as algorithms for tensor environmental and traditional collaborative filtering. Besides LCR, the following baseline algorithms are used in this paper:

\begin{table}
\centering
\caption{Parameters used for each method.}
\begin{tabular}{ccc} \hline
Method & $lastfm$-$50tags$ & $Yelp$ \\ \hline
ILCR &$\alpha$=0.04, $\lambda$=0.01 & $\alpha$=0.08, $\lambda$=0.01 \\ \hline
LCR & $\alpha$=0.04, $\lambda$=0.01 & $\alpha$=0.1, $\lambda$=0.01 \\ \hline
PITF & $\alpha$=0.002, $\lambda$=0.01 & $\alpha$=0.02, $\lambda$=0.01 \\ \hline

\end{tabular}
\end{table}

$\textbf{Pairwise}$ $\textbf{interaction}$ $\textbf{tensor}$ $\textbf{factorization}$ \textbf{(PITF)} \cite{rendle2010pairwise}: As the state-of-the-art tensor algorithm for personalized tag recommendation, PITF is a personalized and context-aware algorithm aiming to return a top-$k$ ranked list of tags when given a particular user-item pair. In PITF, the interaction between each entity pair is expressed as the dot-product of specific feature vectors. The ternary relationship $user$$\times$$item$ $\times$$tag$ is modified by the following score function:

\begin{equation}
    \hat{x}(u, t, i) = \hat{u}_{u}^T\hat{t}_{t}^U + \hat{u}_{u}^{I}\hat{i}_{i}^U + \hat{t}_{t}^I\hat{i}_{i}^T,
    \label{PITF}
\end{equation}
where the first term $\hat{u}_{u}^T\hat{t}_{t}^U$ denotes the relevance value of the given user $u$ and tag $t$, middle term $\hat{u}_{u}^{I}\hat{i}_{i}^U$ denotes the relevance value of the given user $u$ and item $i$, and the last term $\hat{t}_{t}^I\hat{i}_{i}^T$ denotes the interaction between the given item $i$ and tag $t$. According to \cite{rendle2010pairwise}, the user-item interaction term vanishes when $\textsc{LearnBPR}$ is employed to optimize top-$k$ ranking task. Thus, the final score function for tag recommendation task is:

\begin{equation}
    \hat{x}(u, t, i) = \hat{u}_{u}\hat{t}_{t}^U + \hat{i}_{i}\hat{t}_{t}^I.
\end{equation}

Analogously, collaborative retrieval task involves with returning a top-$k$ ranked list of items w.r.t. a given user-query pair, which inspires us to adapt it to CR tasks by lightly modifying Equation (\ref{PITF}). Likewise, the user-query interaction term vanishes, then the modified score function is denoted as:

\begin{equation}
    \hat{x}(u, q, i) = \hat{u}_{u}\hat{i}_{i}^U + \hat{q}_{q}\hat{i}_{i}^Q.
\end{equation}
where the first term $\hat{u}_{u}\hat{i}_{i}^U$ denotes the interaction between user $u$ and item $i$, and the last term denotes the interaction between query $q$ and item $i$. The values of parameters can be leaned by employing BPR algorithm \cite{rendle2010pairwise}.

\textbf{Non-negative Matrix Factorization (NMF)} \cite{lee1999learning}: This approach is not directly applied to model ternary relationship. In this paper, we perform NMF on the item-query matrix to compute a top-$k$ ranked preference items for given $q$ and $u$. The NMF implementation we used in this paper is from http://www.csie.ntu.edu.tw/$\sim$cjlin/nmf/.

In this paper, we conduct numerous experiments to evaluate the performances of each algorithm on the real Last.fm and Yelp datasets. In terms of Last.fm dataset, we randomly draw 80\% of samples in $last.fm$-$50tags$ for training, 10\%  for validation, and the rest for testing. Based on this dataset, we study the efficiency of WARP and BPR on optimizing LCR model (see $\textbf{Section\ 3.4}$). The performance is evaluated on the testing dataset. In addition, we compare ILCR with LCR to validate the capacity of each CR model on retrieving items with different size of training samples (40\%, 60\%, 80\%, 100\% of the total training data). Analogously, we randomly draw 60\% of samples in $Yelp$ for training, 20\% for validation, and the rest for testing. The initial values of feature matrices of both models are randomly drawn from a uniform distribution $\mathcal{U}(-0.02,0.02)$. The hyperparameters $\alpha$ and $\lambda$ are chosen for each algorithm using the validation set respectively. Different experiment settings are used based on the dataset and algorithms (see Table 2).

\subsection{Evaluation}
The performance of each algorithm is measured by recall@$k$, a widely used metric to evaluate the recommendation accuracy in top-$k$. For a given testing example ($q$,$u$,$a$), we first compute $f(q,u,i)$ for each item $i\in \mathcal{A}$, and then sort them in descending order of score. Then recall@$k$ equals to 1 if artist a appears in the top $k$ list, and equals to 0 otherwise. We report mean recall@$k$ over the whole test dataset.

\begin{figure*}
    \centering
    \begin{tabular}{cc}
    \epsfig{file=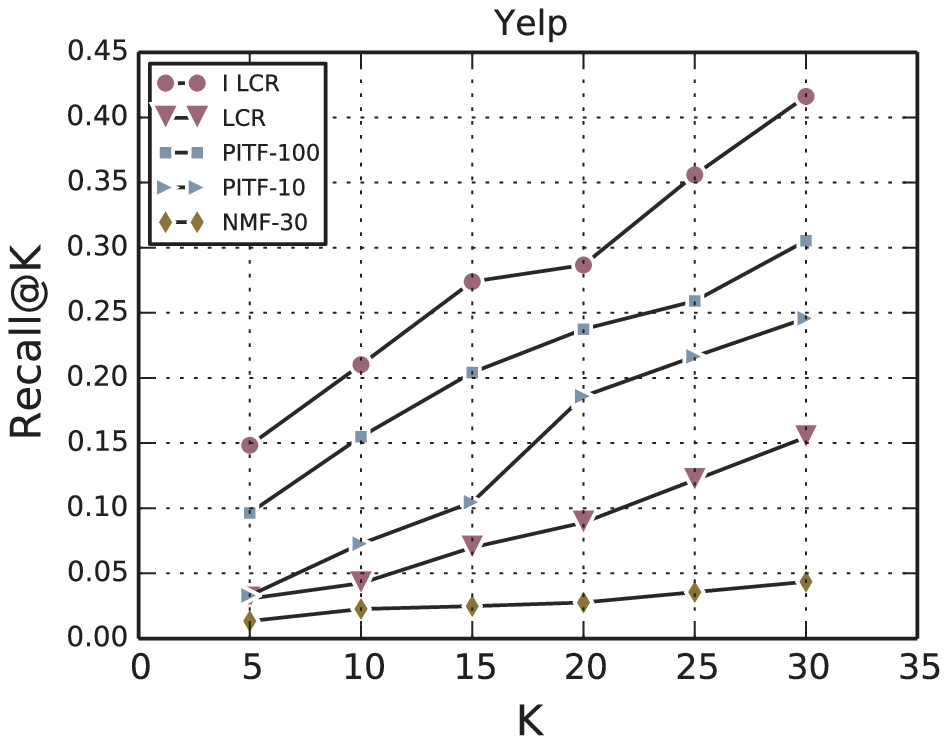, height=210pt, width = 225pt}&
    \epsfig{file=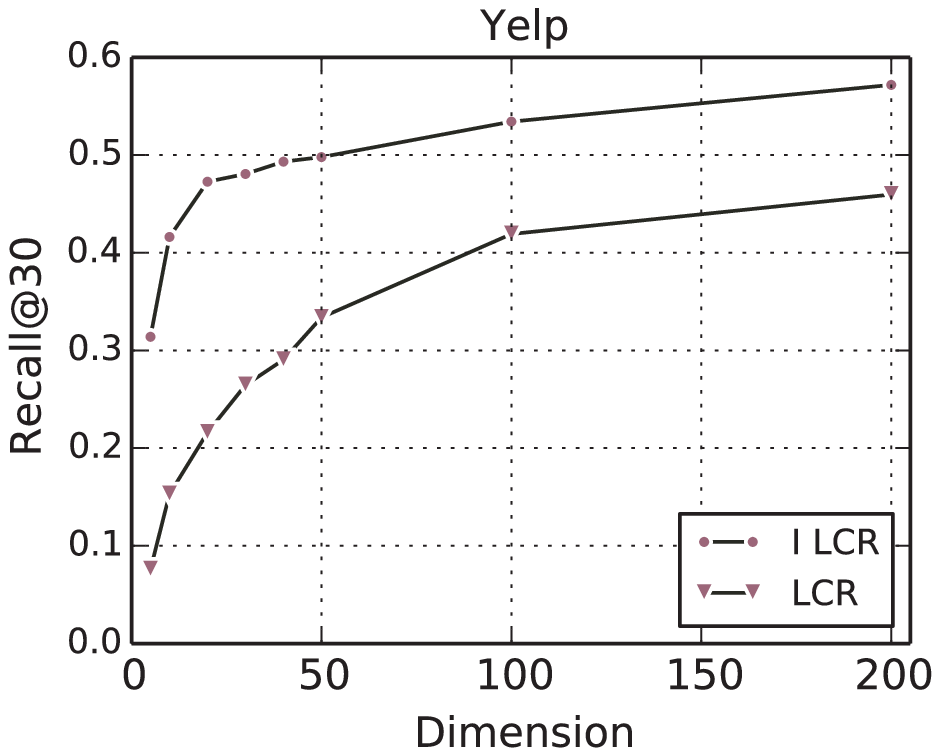, height=210pt, width = 225pt}\\
    (a) & (b)
    \end{tabular}
    \caption{ Experimental results on $Yelp$ dataset. (a) Different algorithms' \emph{Recall} vs. different $k$. The integer numbers of PITF and NMF represent the dimension $n$, and $n$ is set as 10 for ILCR and LCR algorithms. (b) Recall@30 vs. the dimension parameter $n$.}
\end{figure*}

\subsection{Results}
The experimental results on the Last.fm dataset are depicted in Figures 4. From Figure 4.(a), we can see that the correctly predicted items of both ILCR and LCR increase as the growth of different values of $k$ with fixed dimension $n$=10. Comparing with other baseline algorithms, ILCR performs best in retrieving items with given queries by users. Noted that, in comparison with LCR in Figure 4.(a), ILCR increases at least 3.7\% when $k$=30, whilst improves over 12\% in the best case when $k$ equals to 15. In terms of PITF, recall ratio improves along with the increase of both parameters $k$ and dimension $n$. However, PITF still obtains lower recall value than LCR and ILCR on the Last.fm dataset. The performance of NMF suggests that only taking into account binary relationship is insufficient for collaborative retrieval task. Figure 4.(b) shows that the performance of both ILCR and LCR improve along with the increase of dimension $n$, while ILCR outperforms LCR with over 1.9\% improvement when $n$ = 20. In addition to evaluating the efficiency of involved algorithms with different chosen values of $k$ and $n$, the performance of ILCR and LCR on various size of training samples is also shown in Figure 4.(c). The sparsity of training dataset declines as the growth of the size of training samples. Figure 4.(c) reveals that ILCR could effectively represent the latent relationship of the ($user$,$query$,$item$) triples, even for different sparse datasets.

Since we expect the integration of item-based information to be particularly useful when the dataset includes massive amount of sparse items, we also show experimental results on a much sparser dataset, $Yelp$. Figure 5.(a) shows that the recall value of both ILCR and LCR improve with the increase of parameter $k$. From Figure 5.(a) and (b), we can see that the performance of LCR on the $Yelp$ dataset significantly decreases with different settings, in contrast with the well performance on the $Last.fm$ dataset. It's noted that we carefully and repeatedly choose the best setting of key hyperparameters $\alpha$ and $\lambda$ on the validation dataset. However, the difference of recall ratio between LCR and ILCR is still evident, which suggests that ILCR can be useful for evaluating the ranks of items, especially those sparse items. Figure 5.(a) shows that ILCR outperforms the selected baseline algorithms at different values of $k$ with fixed dimension parameter $n$=10. The experimental results demonstrate that taking item's collaborative information into account can effectively predict users' preferences on items w.r.t. a given query under a situation where there is imperfect knowledge of the characteristics of the majority proportion of items.


\section{Related Work}
Recommendation and retrieval techniques have become essential components of massive applications, like E-commerce, search engine, music-based social network etc. In information retrieval, one is expected to rank items or documents by using the content features of items depending on a given query by an active user. In that case, many algorithms are proposed, like Latent Semantic Indexing \cite{Deerwester90indexingby}, LDA \cite{Blei03latentdirichlet} topic model for low-dimensional representation of the word. More recently, factorized models like Polynomial Semantic Indexing (PSI) \cite{bai:polynomial} are employed to implement the task of document retrieval. Methods like PSI for retrieval task optimizes the AUC ranking loss, which does not optimise the top $k$ ranked list like ours. In recommendation field, many works based on factorized models are proposed to produce predictions to active users. In particular, Matrix Factorization-based methods \cite{koren:matrix, koren:factorization, PMF} are very close to our method since each entity (e.g. user, item, query) is represented as a low-dimensional feature vector. Main difference between ILCR and these general matrix factorization methods is that each recommendation in CR task is seeded with a query, while most of them are invented to deal with binary user-item relationship. In fact, the most related works to our model are those models which can be adaptive to deal with multi-types of relationship as ternary interactions are considered in CR task. There are several available choices, such as classical Tucker decomposition \cite{1982489}, PARAFAC. Many context-aware collaborative filtering techniques are also proposed for such multi-relationship learning task, in particular contextual information related to users, like tags \cite{rendle2010pairwise}, web pages  \cite{krishnamenon:response}, demographics \cite{karatzoglou:multiverse}, mobile information \cite{13758559}, temporal effects \cite{13270124}. In addition, recent works on heterogeneous information network \cite{sun:pathsim, yu:personalized} can also be adaptive to give vision into the sophisticated interactions hidden in the complicated dataset.

\section{Conclusions and Future Works}
We aim to design an effective collaborative retrieval algorithm to objectively predict the ranks of those items with lack of descriptive terms revealing the basic characteristics of them. To achieve this goal, we focus on the profits of leverage the collaborative information of items to better evaluate the ranks of those sparse items, and propose to express the latent sophisticated relationships for CR task from not only users' perspective, but also items' perspective. Then, we propose a superior latent collaborative retrieval model, ILCR, by integrating the possible item-based information into LCR model. Finally, we study the primary principles of LCR, of which we find that the runtime of LCR is dominated by the chosen learning approach, namely WARP.  Therefore, we propose to employ a generic approach, namely BPR, instead of WARP to optimize the LCR model on the basis of further experimental analysis covering the topics of training efficiency and accuracy. 

The proposed model can be easily generalized to deal with many other tasks involving to model ternary interaction among entities, not limited to CR task. Most typical example might be the tagging recommendation system, which aims to recommend tags to users w.r.t. items. However, we just explore possible interactions among $(user, query, item)$ triple. In practice, relationships of a pair of entities always involve with massive ingredients, which could  be termed as a currently-prevalent word,  heterogeneous relationships. Due to this significant characteristics, we will attempt to make our proposed model adaptive to learn the multi-relation by exploring diverse aspects of a sophisticated system in future work.

\section{Acknowledgements}
This work was partially supported by the National Natural Science Foundation of China (No. 11305043), and the Zhejiang Provincial Natural Science Foundation of China (No. LY14A050001), the EU FP7 Grant 611272 (project GROWTHCOM) and Zhejiang Provincial Qianjiang Talents Project (Grant No. QJC1302001).

%
\bibliographystyle{abbrv}
\bibliography{recsys}  
%
%
\end{document}